\documentstyle[prl,aps,multicol,epsfig,amsmath,amssymb]{revtex}
\begin{document}
\draft

\title{Interplay between geometry and flow distribution in an airway tree}
\author{B. Mauroy$^1$, M. Filoche$^2$, J. S. Andrade Jr.$^{1,3}$,
and B. Sapoval$^{1,2}$}

\address{
$^1$Centre de Math\'ematiques et de leurs Applications, CNRS, Ecole Normale
Sup\'erieure de Cachan, 94235 Cachan, France\\
$^2$Laboratoire de Physique de la Mati\`ere Condens\'ee, CNRS, Ecole
Polytechnique,
91128 Palaiseau, France\\
$^3$Departamento de F\'\i sica, Universidade Federal do Cear\'a,
60451-970 Fortaleza, Cear\'a, Brazil}

\date{\today}

\maketitle

\begin{abstract}
Uniform flow distribution in a symmetric volume can be realized
through a symmetric branched tree. It is shown here however, by 3D
numerical simulation of the Navier-Stokes equations, that the flow
partitioning can be highly sensitive to deviations from exact symmetry
if inertial effects are present. The flow asymmetry is quantified and
found to depend on the Reynolds number. Moreover, for a given Reynolds
number, we show that the flow distribution depends on the aspect ratio
of the branching elements as well as their angular arrangement. Our
results indicate that physiological variability should be severely
restricted in order to ensure adequate fluid distribution through a
tree.
\end{abstract}


\maketitle

\begin{multicols}{2}
The problem of fluid flow in a branching geometry appears in many
physical, geological, chemical and biological systems. Examples
include catalysis, flow through porous media, blood circulation and
respiration. When studying transport in any of these systems, a common
objective is to understand the mechanisms that govern the flow
partitioning at the interconnections level. Until recently, it has
been generally assumed that the use of Darcy's law should be
sufficient to describe the propagation of flow through branched
structures. Such a relation corresponds to the linear dependence
between flow and pressure drop, $Q \propto \Delta P$, which is
strictly valid at small Reynolds number. Regardless of this
limitation, a large number of studies have been based on this
approximation. In the context of porous media, for instance, a simple
paradigm to represent flow through the pore space is a network of
bifurcating and merging channels where the transport of fluid is
analogous to the distribution of electrical currents in a resistor
network. However these models can only predict a perfectly uniform
and synchronous flow distribution through airways bifurcations
\cite{Ped70}. A major problem in modeling of flow through
trees arises from the fact that, due to inertial effects, Darcy's law
breaks down as a phenomenological description for large Reynolds
numbers. Even at moderate Reynolds, the inertial non-linearities become
relevant as compared to viscous effects.

Unambiguous experimental and numerical evidences of inertial effects
have been observed in several studies on flow though branched
structures, with special emphasis on the bronchial tree
\cite{Gro01,Sny81,Sny83,Slu80,Isa82,All85,Tsu90,Wil97,And98,Zha97,Mar01,Com01,Cha89}.
Such phenomena exists in real lungs but they are more
simple to study in a symmetric geometry \cite{Man82,New97}. In
particular, in order to irrigate uniformly a symmetric volume it is
easy to show, through the following collage argument, that this is
ensured by a symmetric tree. Suppose that an asymmetric tree feeds a
volume which has a plane of symmetry. If the tree is asymmetric the
flow will be different in the two parts of the volume which are
symmetrical. Then one can replace the tree with a non-uniform flow
by the symmetry image of the more efficient region. The new tree,
which is now symmetric, is more efficient for flow distribution.

In the Poiseuille approximation, the only way to have perfect symmetry
is to work with an equivalent resistor network that is symmetric. In
other words, at each bifurcation the daughter branches should be
exactly identical irrespectively of their real geometrical
arrangement. This might not be true if inertial effects are
present. It should be recalled that, as the lung is a succession of
branch bifurcations, the final flow distribution can be represented by
a multiplicative process. In consequence, even a rather small
asymmetry could lead to a strong inhomogeneity of the flow
distribution \cite{Sap97}. Because the geometrical arrangement of the
bronchial tree of mammals is always subjected to some physiological
variability \cite{Shl91}, it appears natural to question whether a
small modification of the structure disturbs the distribution of fluid
flow.

\begin{figure}
\centering
\includegraphics[width=6cm]{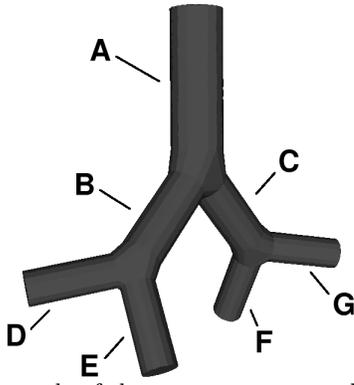}
\caption{Example of the tree geometry used in the simulations.
The aspect ratio is $L/D=3$ and the rotation angle is $\alpha=45^{\circ}$.}
\label{tree}
\end{figure}

\begin{figure}
\centering
\includegraphics[width=6cm]{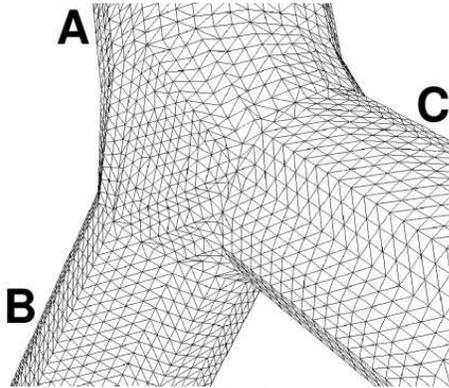}
\caption{Geometry and mesh of a typical bifurcation used in the
simulations.}
\label{mesh}
\end{figure}

The purpose of this work is to investigate how the tree geometry
influences the flow in order to shed some light on the optimal aspect
of the bronchial tree for distributing air uniformly in the lung
volume. The direct 3D numerical solution of the Navier-Stokes
equations is by far the most practical way to elucidate this
problem. The simplified tree model used here is shown in
Fig.~\ref{tree}. It consists in a 3-dimensional cascade of cylinders
branching through two bifurcations. Each bifurcation ABC or BDE or
CFG, is coplanar as found approximately in real lungs. The bifurcation
geometries are modeled in such a way to minimize geometrical
singularities as shown in Fig.~\ref{mesh}. For simplicity, we assume
that the radii of the tubes decrease with a factor $2^{-1/3}$ at each
bifurcation \cite{Wei84} and choose the branching angle to be
$45^{\circ}$.

The mathematical description for the detailed fluid mechanics in the
branched structure is based on the steady-state form of the continuity
and Navier-Stokes equations for mass and momentum conservation \cite{flow}
\begin{equation}
{\bf{\nabla\cdot u}}=0~, \label{eq1}
\end{equation}
\begin{equation} \rho~{\bf{u\cdot\nabla u}} = -{\nabla p} +
\mu~{\bf{\nabla}}^{2}{\bf{u}}~, \label{eq2}
\end{equation}
where {\bf u} and $p$ are the local velocity and pressure fields,
respectively. The diameter of the first tube is equal to $2$~cm,
corresponding approximately to the diameter of the human trachea.
The fluid is air with viscosity $\mu = 1.785 \times 10^{-5}$~
$\rm{kg~m^{-1}s^{-1}}$ and density $\rho=1.18$~$\rm{kg~m^{-3}}$, and
the flow is considered to be incompressible. Nonslip boundary
conditions are imposed at the tube walls (Dirichlet condition ${\bf
u}=0$) and the velocity at the entrance A is parabolic. The outlets
are free with the same reference pressure and $\partial {\bf u}/ \partial
n=0$.
Equations~(\ref{eq1}) and
(\ref{eq2}) are solved using finite elements \cite{detail}.
For all simulations, the relative conservation error is
smaller than $3\%$.

The parameters governing the flow are the bronchi aspect ratio (length
to diameter ratio of the tubes) $L/D$, the rotation angle $\alpha$
between successive bifurcations, and the Reynolds number, $Re \equiv
\rho D V/\mu$, where $V$ is the mean velocity at the entrance. The
reference angle $\alpha = 0^{\circ}$ corresponds to a coplanar tree.
The flow asymmetry is defined as
\begin{equation}
\Sigma(\alpha, L/D) \equiv \left|\frac{q_1-q_2}{q_1+q_2}\right|~,
\end{equation}
where $q_1$ and $q_2$ are the outflows at (D,G) and (E,F) branches,
respectively. We perform simulations for several values of $\alpha$,
$L/D$ and $Re$ to find their influence on the flow partitioning
$\Sigma$. Note that the air velocity at the entrance of human lungs
range from $1$~m/s at rest ($Re \approx 1200$) to $10$~m/s for the
condition of very hard exercise ($Re \approx 12000$) \cite{Wei84}. Due
to the number of parameters governing the flow and the computation
time for each set of parameters, we first discuss the dependence of
the flow asymmetry on the geometry for a fixed Reynolds value, namely,
$Re=1200$. This corresponds to the human inspiration state at rest.

\begin{figure}
\centering
\epsfig{file=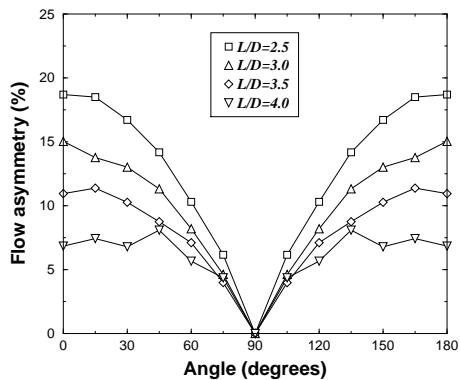,width=6cm}
\caption {Dependence of the flow asymmetry $\Sigma$ on the branching
angle $\alpha$ for a fixed Reynolds value, $Re=1200$. The observed
non-monotonous dependences are due to numerical uncertainties.
The values of $\alpha=0^{\circ}$ and $180^{\circ}$ correspond to
a planar tree. $\alpha = 90^{\circ}$ represents the average value
for mammalian lungs.}
\label{angle}
\end{figure}

The results are shown in Fig.~\ref{angle}. The main result is that,
whatever the conditions, the behavior of $\Sigma$ around the minimum is {\it
not parabolic}. Even a small departure from geometrical symmetry can
cause a non negligible flow disturbance. For a given value of $L/D$,
the disturbance increases with the deviation of the rotation angle
from $90^{\circ}$. $\Sigma$ is therefore maximum for a planar tree and,
for a fixed $\alpha$ value, it decreases with increasing aspect ratio.

\begin{figure}
\centering
\epsfig{file=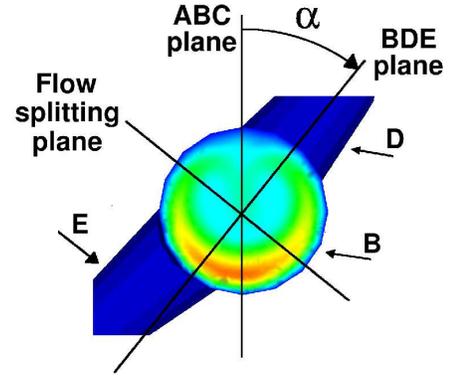,width=7cm}
\caption{The M-shape contour for $L/D=3$ and $\alpha=45^{\circ}$.
The colours indicate the magnitude of the fluid velocity at the
mid-length cross-section of branch B. The velocity magnitude increases
in the colour order of blue, green, yellow and red. The ternary branches D and
E are shown in blue. Note the presence of a low velocity region at the
center. At the plane of the second bifurcation, the entering flow is
larger at the bottom. The branch E therefore captures a larger flow than D.}
\label{mshape}
\end{figure}

There are then two facts to interpret. First, why the flow is
influenced by breaking the symmetry only. Seconds why this effect is
attenuated for long branches or large aspect ratios. The first fact
can be understood by considering the velocity distribution in a cut of
the secondary branch B shown in Fig.~\ref{mshape}. The flow keeps the
symmetry of the ABC bifurcation plane but, due to inertia, the high
velocity regions are drifted vertically and an M-shape type of
distribution is observed \cite{Wil97}. This shape governs the flow
partitioning at the second bifurcation. Note that if the branches B
and C are long enough and for small $Re$, the profile should tend to a
parabolic type. As a consequence, the distribution shown in
Fig.~\ref{angle} will progressively change along the second generation
branch. It is because the branch length is too short that the
grand-daughter branches can capture the asymmetry seen in
Fig.~\ref{mshape}. This provides a qualitative answer for the second
question. The position of the intersection relatively to the M-shape
is then the key for asymmetry. For example in Fig.~\ref{mshape}, the
branch E obviously receives more flow than branch D. It is also clear
that, if $\alpha=90^{\circ}$, the flow symmetry is restored for any
value of $L/D$.

\begin{figure}
\centering
\epsfig{file=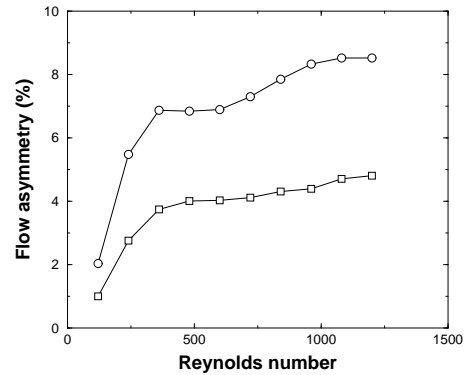,width=6cm}
\caption{Dependence of the flow asymmetry $\Sigma$ on the Reynolds
number $Re$ for $L/D=3$. The circles
correspond to $\alpha=60^{\circ}$ and the squares to $\alpha=75^{\circ}$.}
\label{reynolds}
\end{figure}

\begin{figure}
\centering
\includegraphics[height=6cm]{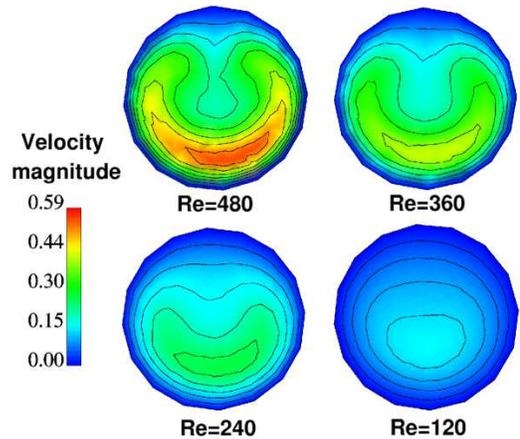}
\caption{Contour plot of the velocity magnitude at the cross-section
of the second bifurcation for different values of $Re$ ($L/D=3$ and
$\alpha=60^{\circ}$). As $Re$ increases, the profiles gradually change
from parabolic to M-shape.}
\label{contours}
\end{figure}

The dependence of the flow asymmetry as a function of the Reynolds
number is shown in Fig.~\ref{reynolds}. A strong increase of $\Sigma$
is observed up to $Re \approx 250$ followed by a region of weaker
dependence. This type of behavior has been previously reported for 2D
flow in trees comprising more than two generations of branches
\cite{Wil97,And98}. It is remarkable that the onset at $Re \approx
250$ is approximately the {\it same} whatever the angle $\alpha$. This
is a clear indication that, at the entrance of the second bifurcation,
the velocity profile reaches the same pattern for a given $Re$
value. Again, the final asymmetry of the distribution of flow is a
result of the inertial effects originated from the first
bifurcation. All these arguments are illustrated in Fig.~\ref{contours},
where the contour plots of the velocity fields are shown at the
entrance of the second bifurcation. At large $Re$, the M-shape is
revealed and, as expected, the lower the $Re$, the closer the profile
is to parabolic flow. The smaller variation of $\Sigma$ for $Re > 250$
can be explained by the presence of a secondary flow \cite{Com01}.

Some implications of our results are noteworthy. If the inertial
effects observed here are present in a larger tree, the relative flows
delivered to the outlets of this structure may become strongly
non-uniform. This broadness in the flow distribution is a typical
signature of a multiplicative process \cite{Wes89}, where an
observable can be viewed as a ``grand process'' depending on the
successful completion of a number $n$ of independent
``subprocesses". It is then possible to associate the flow at each
branch with a probability $p_i$, so that the flow at a given outlet
$k$ be $q_{k} \propto p_1p_2 \cdots p_n$, where $i=1,2,3,\cdots,n$
corresponds to the set of branches constituting the pathway going from
the entrance to the exit $k$. It can be easily demonstrated that, if
the $p_i$'s are independent variables and $n$ large, the distribution
of $q_{k}$ should be approximately log-normal. Furthermore, this
distribution might {\it mimic} a power-law if its dispersion is
sufficiently large \cite{Wes89}. Note that this situation is that of
the human bronchial tree (where $L/D$ is close to $3$) even at
rest. In this case $Re \approx 1200$ and the multiplicative process
due to inertia can propagate further down in the tree. If we consider
that these effects only disappear for $Re$ less than 100 and that the
local Reynolds decreases by a factor of $2^{2/3}$ at each generation,
we obtain that the flow asymmetry can be significant up to the 6th
generation of the bronchial tree under rest conditions \cite{Suki94}.

In conclusion, we have investigated the effect of inertia on fluid
flow through three-dimensional rigid branched structures by direct
numerical simulation of the Navier-Stokes equations. It has been found
that for trees with $3$ generations of cylindrical conduits, the flow
distribution at the outlets strongly depends on the Reynolds number
and on the geometry of the ramified structure. Moreover, our
simulations indicate that the flow imbalance throughout the tree is
highly sensitive to the aspect ratio $L/D$ of its cylindrical units
and to the variation of the rotation angle $\alpha$ between
successive bifurcations. While a uniform distribution of flows at the
outlets of the third generation branches is always obtained for
$\alpha=90^{\circ}$, our calculations show that a small deviation from
this geometrical configuration is capable to induce a large asymmetry
on the flow. Note that the presence of long branches would lead to
purely axi-symmetric parabolic profiles and flow symmetry. However,
long tubes exhibit large hydrodynamic resistance (proportional to
$L/D^4$). It is therefore not surprising that in real lungs $L/D
\approx 3$ and $\alpha \approx 90^{\circ}$.

Finally, our results suggest that small deviations from the ``best''
structure should have the same type of consequences in the real
(asymmetric) lung, namely, strong dependence on geometry and Reynolds
number. In particular, the flow distribution at rest and exercise
might be significantly different. These results could also help to better
understand
lung morphology.
It has been argued \cite{Nel88} that the asymmetric structure of the lung
is solely due to geometrical constraints. Our study indicates that the
inertial effects plays also an important role in air distribution. In other
words, the asymmetry of the
bronchial tree is determined not only by geometrical constraints but also
by the existence of inertial effects. Of course, if the flow
distribution is found uniform although the geometry is ``imperfect'', the
following question would naturally arise \cite{Bec01}: what are the
physiological regulation mechanisms that can compensate the flow non-uniformity
due to inertial effects? In addition, the
fluid dynamics studied here is certainly relevant to understand
particle deposition in the airway tree \cite{Com01}, a problem of
crucial importance both from the physiologic and the therapeutic
points of view.

We thank CNPq, CAPES, COFECUB and FUNCAP for support. The Centre de
Math\'ematiques et de leurs Applications and the Laboratoire de
Physique de la Mati\`{e}re Condens\'{e}e are ``Unit\'{e} Mixte de
Recherches du Centre National de la Recherche Scientifique'' no. 8536
and 7643.

\end{multicols}

\end{document}